\documentclass[12pt]{article} 
\usepackage{array,epsfig,fancyheadings,rotating}
\usepackage[]{hyperref} 

\usepackage{amsmath}
\usepackage{amssymb}
\usepackage{amsfonts}
\usepackage{multirow}
\usepackage{amsthm}

\theoremstyle{definition}

\title{Case based error variance corrected estimation of structural models}
\author{Reinhard Oldenburg, Augsburg University, Augsburg, Germany}
\date{4th October 2021}

\begin{document}
	
	\maketitle
	
	\begin{abstract}
	A new method for estimating structural equation models (SEM) is proposed and evaluated. In contrast to  most other methods, it is based directly on the data, not on the covariance matrix of the data.   The new approach is flexible enough to handle non-linear and non-smooth models and allows to model various constraints. Principle strengths and weaknesses of this approach are discussed and simulation studies are performed to reveal problems and potentials of this approach.
		\end{abstract}

	\def\thefigure{\arabic{figure}}
	\def\thetable{\arabic{table}}
	
	\renewcommand{\theequation}{\thesection.\arabic{equation}}

\section{Introduction}

It is needless to say anything about the importance of structural equational modeling (SEM). A good overview of its
development and application is given in \cite{Hoyle2012}. However, despite the many virtues of SEM and the many variations that have been developed over time, there are a number of reasons  to look for ways to extend the framework and gain more flexibility in modeling. 
This paper will introduce and explore a very elementary least squares
estimation procedure which is, however, computationally demanding. More than 40 years ago, when SEM first took shape, computers were simply not fast enough to perform these calculations. Hence, back then it was a clever step not to work directly with the data and the model equations. Instead, the so called LISREL approach \cite{Bollen1989} is characterized by calculating a parameter-dependent covariance matrix from the model equations and fitting this to the empirical covariance matrix of the data. With modern computers, however, fitting the model equations directly became feasible, and hence it is worth investigating the benefits (and the drawbacks) of this approach. 

To explain the characteristics of the proposed method  a quick review of the classical approach to SEM (\cite[p. 319]{Bollen1989}) is needed to make the difference clear.  The traditional approach assumes that there are measurement models for  exogenous $\xi$  and endogenous $\eta$  latent variables in terms of observed variables  $x,y$:
\begin{equation}
	\begin{split}
		x=\Lambda_x \xi +\delta \\ y=\Lambda_y \eta +\epsilon
	\end{split}
\end{equation}
All these are vectors of random variables with real coefficient matrices $\Lambda_x, \Lambda_y$ and vectors of error  variables $\delta,\epsilon$.

The structural model is given by the linear relation between the latent variables and it involves further structural matrices $B,\Gamma$ and another error term  $\zeta$: 
\begin{equation} \eta =B\eta +{\Gamma} \xi +\zeta\end{equation}
Assuming that $I-B$  is non-singular (with $I$ the identity matrix of appropriate size) one has the equation  
$\eta =\left(I-B\right)^{-1}(\Gamma\xi +\zeta )$ that allows to eliminate the endogenous latent variables.  Moreover, one needs the assumptions (this can be relaxed to allow a limited number of violations)
\begin{equation}\label{errassume}
	\mathrm{cov}\left(\xi ,\delta \right)=0, \mathrm{cov}\left(\xi ,\epsilon
	\right)=0, \mathrm{cov}\left(\zeta ,\delta \right)=0, \mathrm{cov}\left(\zeta ,\epsilon \right)=0
\end{equation}
Under these assumptions the parameter implied covariance matrix can be calculated:
\begin{equation}
	\Sigma=\left(\begin{array}{rr}\mathrm{cov}(x,x')&\mathrm{cov}(x,y')\\
		\mathrm{cov}(y,x')&\mathrm{cov}(y,y')
	\end{array}\right)
\end{equation}

In the end, the model shall describe some data, i.e. one needs $n$ realizations of the observed variables $x_i,y_i, i=1\ldots n$ and from this one can calculate the empirical numerical covariance matrix $S$. The estimation step is then to minimize some distance measure between $S$ and $\Sigma$. The minimizer is sought among all free parameters from $B,\Gamma,\Lambda_x,\Lambda_y$ as well as the unknown covariances between the exogenous latent variables and the covariance matrix of the error vectors $\delta,\epsilon$. To make the model identified, usually many of these error correlations must be assumed to be zero (and, of course, this is in  general a sensible assumption for error variables). 

To sum up the characteristics of this approach: The problem size is reduced by eliminating latent variables and keeping only covariances between some of them. To do this, a lot of assumptions are needed already on the model level, some further for specific estimation procedures.

In contrast the present paper evaluates the possibility of an approach that takes the errors of the model equations in each individual case of the data set as the central objects. For the linear SEM model described above these errors are: 
\begin{equation}
	\begin{split}
		\delta_i=x_i-\Lambda_x\xi_i  \\ 
		\epsilon_i=y_i-\Lambda_y\eta_i \\
		\zeta_i=\eta -B\eta_i -\Gamma \xi_i 
	\end{split}
\end{equation}
The approach then is to minimize a weighted sum of the errors 
\begin{equation}
	\textrm{min}\sum_{i=1}^n w_\delta\cdot\delta_i^2+w_\epsilon\cdot\epsilon_i^2+w_\zeta\cdot\zeta_i^2
\end{equation}
The choice of weights is non-trivial and, of course, some strategy is needed to fix them and this will be discussed. The minimizer gives estimates both for the coefficients from $B,\Gamma,\Lambda_x,\Lambda_y$ as well as the latent variables for each case $i$. From this information, individual residuals can be calculated. The price to pay is of course that a much more demanding optimization problem has to be solved. But it will be argued, that this can be done for sensible sample sizes. The benefits are obviously that much less assumptions are needed and hence a wider class of models can be estimated, including non-linear models. Below the general framework will be described and applied in several simulation studies. 

The rest of this introduction relates the main characteristics of the proposed method, namely estimation of latent variables, moderate assumptions and extensions to nonlinear models, to the literature.

Estimates for latent variables can be obtained in various ways, e.g. factor scores and composites. Factor score methods (e.g. regression scores and Bartlett scores, see \cite{DZM2009} and \cite{Yung2013})  first estimate a SEM model and then use this information to arrive at estimates for the latent variables,  the factor scores. Obviously, this limits the determination of scores to that cases for which SEM can be calculated. Moreover, this separates the estimation process into two steps while it might seem better to have one coherent estimation step.  Composite methods in a sense go the opposite direction, i.e. they first determine composites as weighted sums of manifest variables and substitute them for the latent variables. The debate if these composites can be viewed as latent variables is subject to discussion (e.g. \cite{RE2013}, \cite{HDSR2014}).  A popular composite method is that of partial least squares PLS (\cite{Hwang2017}  and  \cite{ECH2010} provide an overview).  The composites are usually seen as proxies for the latent variables, reflecting especially that in composites, indicators don't have individual error terms. 
Another issue is highlighted by  \cite{Hwang2017}.  They criticize  that in PLS there is no coherent optimization goal that determines the solution. Hence, it is difficult to come up with easy to interpret measures of model fit. As a solution they propose a method called general structure component analysis (GSCA). In this approach there is a linear measurement model (indicators that measure latent variables) and a separate linear structural model between latent variables. Moreover, there is a component model that determines how to calculate estimates for latent variables as linear combinations of manifest variables   \cite[p. 18]{Hwang2017}. GSCA has been extended to include some measurement error equations \cite{HTK2017} and to allow the modeling of common factors \cite{HCJ}. However, these approaches incorporate two models (measurement and composite) in one leading to a double determination of the latent variable which obscures somewhat the meaning of the estimates. 

Bayesian variants of SEM, e.g. as realized in the blavaan software package \cite{ME2015}, are more flexible and give estimates for latent variables. However, the computational demands are typically much higher than those of the method proposed in this paper.

As described above, in the standard approach to SEM one passes from an analysis of the data to an analysis of its covariance matrix. This step from the model equations to $\Sigma$ rests on some assumptions that may be violated by real-world data, e.g.  one must assume vanishing covariance between latent and error variables and, moreover, the number of residual covariances between error variables is limited by the need of identification of the model.  However, if the goal is simply to test if a model without restrictions on the residual covariances  would fit better than the model at hand, then the Lagrange multiplier test (see \cite[p. 293]{Bollen1989}) gives an answer. However, it does not give estimates of error covariances and it is not always applicable (it has the most power for small deviations as it is based on the second order Taylor expansion of the likelihood function; and, it is asymptotic).    More widespread is the concept of modification indices that can be used to test if certain correlations of error variables are severe \cite[p. 372]{Hoyle2012}, but in this process only the effect of relaxing one constraint can be estimated, not all of them simultaneously. 

Originally SEM  dealt only with linear models. There are, however,  many extensions to non-linear models (see e.g. \cite{SM1998}  and \cite{UNBK2017}), but many of them lack the ease of use that linear models have. Some approaches are limited to very special generalizations, e.g. quadratic terms. While this has the advantage that some distribution theoretic results can be achieved  (e.g. \cite{KWS2011}), it could mean that the modeling flexibility is insufficient to match certain situations, e.g. piece-wise linear models cannot be treated that way. Moreover, PLS and GSCA can be extended to some nonlinear models as well (e.g. \cite{DS2014} for PLS and  \cite[ch. 5]{HTK2017}  for GSCA). 
However, it seems that there is no approach that can handle  piece-wise functions like 
$$y=\left\{\begin{array}{lr}a{\cdot}x+b & x<x_0\\
	a{\cdot}x_0+b+c{\cdot}\left(x-x_0\right) & x {\geq} x_0\end{array}\right.  $$ 
Being able to model with them would enhance the benefit of SEM by allowing to identify points ($x_0$) where the slope of the linear relation
changes, e.g. beyond a certain extent an influence may no longer enhance some result. Similarly, relations between two
(centered latent) variables $x,y$ of the form 
$\Theta \left(x\right){\cdot}\left(y-k{\cdot}x\right)=0$, where  $\Theta$  is the Heaviside function, model implicative relations as will be shown in an example below.   
Non-linearity is often associated with growth curve models 
(e.g. \cite[p. 532]{Hoyle2012}), but in these models, the relations between latent variables remain linear. An approach that may be flexible enough to handle piece-wise and nonlinear models uses splines \cite{Guo2012}  but it seems that there is no easily accessible implementation.  

Maximum likelihood estimation in SEM requires the covariance matrix of observed data to be non-singular. This may seem natural, but consider a reflexive measurement model with two or more observed variables measuring the same latent variable.  If two measurements are almost perfect so that they produce (almost) the same results, the covariance matrix will be (almost) singular. Thus, improving the measurement procedure can render a perfect model to be out of reach of the estimation procedure.  

When the assumptions of SEM are violated, results can be very misleading. The demanding assumptions made in SEM have been one reason for Wold to develop the PLS approach, according to \cite[p. 24]{ECH2010}, and these assumptions are  often violated in real world data as reported in \cite[p. 186]{HDSR2014}. Kline \cite{Kline2012} reviews SEM assumptions and doubts that most researchers know about all of them. 

As mentioned above, a point that distinguishes the proposed method CLSSEM to the methods mentioned above is that CLSSEM is completely rooted in the data. Taking the observed data directly and using the proposed model in its original form seems to be the most natural approach because it suggests that the modeled meaning of construct fits the intended meaning. Similar data centered approaches have not been investigated for SEM but have been studied in the context of exploratory factor analysis (\cite{Unkel2010}, \cite{Adachi2012}, \cite{Adaci2020}). 

The method presented in this paper addresses all of these points. Of course, as is to be expected, the method also has some issues and drawbacks, which will be discussed later.  In evaluating the new method it will be compared to some established methods and available implementations. 

To sum up, the CLSSEM approach presented in this paper is characterized by being rooted in the data and taking the model equations directly as the objects of fitting. It does neither require special assumptions about distributions nor is it restricted in the type of equations used for modeling.

\section{The theory of case based least square SEM (CLSSEM)}\label{sec:2}
The traditional framework of SEM eliminates the latent variables from the objective function that is subject to optimization  by passing to the covariance matrix and replacing the individual values of latent variables by their correlations and variances of associated error variables. The key  idea of the present approach is rather not to eliminate the latent variable but to estimate their values.

The data that is analysed by the methods described in the paper consist of a numeric matrix 
\begin{equation}\left(A_{i,j}\right),i{\in}\left\{1,{\dots},n\right\},j{\in}\left\{1,{\dots},k\right\},A_{i,j}{\in}\mathbb{R}
	\end{equation}
	  of  $n$  cases
for which  $k$  measurements have been performed. These measurements are denoted by manifest variables 
$x_1,{\dots},x_k$. Hence, the vector  $A_{\cdot ,j}{\in}\mathbb{R}^n$  contains the  $n$  measurements of  $x_j$. 
The equational model that is to be fitted to this data consists of a set of  $m$  equations 
$g_l\left(\left\{x_j\right\},\left\{\eta_q\right\},\{p_s\}\right)=0,l{\in}\{1,{\dots},m\}$  that relate the measured
variables with latent variables  $\eta_q$, $q{\in}\{1,..,Q\}$  and parameters  $p_s{\in}\mathbb{R},s{\in}\{1,..,S\}$. Latent variables
are thought to depend on the case just like the observed variables, i.e. for each case there is a hypothetical value 
$Z_{i,q}{\in}\mathbb{R}$  of $\eta_q$ while the parameters are case independent properties of the model. The equations may be linear or non-linear and may also include inequalities (note that the inequality  $h{\geq}0$  is equivalent to  $f(h)=0$  for the
function  $f(x)=x-|x|$).

The instantiation of an equation  $g_l\left(\left\{x_j\right\},\left\{\eta_q\right\},\{p_s\}\right)=0$  for case 
$1\leq i \leq n$  is the equation  $g_l\left(\left\{A_{i,j}\right\},\left\{Z_{i,q}\right\},\{p_s\}\right)=0$ 
and due to measurement errors and due to possible differences between the model and the \textit{true} structure, these
equations are expected to hold only with an error, i.e. 
\begin{equation}\label{geq}g_l\left(\left\{A_{i,j}\right\},\left\{Z_{i,q}\right\},\{p_s\}\right)=\epsilon _{i,l}	\end{equation}  where the error vector
variables  $\epsilon _l$  are required to have mean 0 and a variance as small as possible. The aim of estimating the model is to determine the  $n\cdot Q$  numbers  $Z_{i,q}$  and the  $S$  parameters  $p_s$  so that the errors are as small as possible. Hence, following the least square approximation idea, the goal is to minimize the objective function 

\begin{equation}\label{objfun}
	F_w\left(\left\{Z_{i,q}\right\},\{p_s\}\right):=
	 \sum _{l=1}^m w_l\cdot \sum _{i=1}^n \epsilon _{l,i}^2=
	 \sum_{l=1}^m w_l\cdot \sum _{i=1}^n\left(g_l\left(\left\{A_{i,j}\right\},\left\{Z_{i,q}\right\},\{p_s\}\right)\right)^2
\end{equation}
Here $w\in\mathbb{R}^m$ is a vector of positive real numbers (weights) of the equations.  
Furthermore, in the optimization process one may wish to put some further constraints:
\begin{itemize}
	\item All (or some) latent variables shall be centred: \newline ${\forall}q{\in}\left\{1,{\dots},Q\right\}:\sum _{i=1}^n Z_{i,q}=0$ .
	\item All (or some) latent variables shall be normalized: \newline ${\forall}q{\in}\left\{1,{\dots},Q\right\}:\sum _{i=1}^n Z_{i,q}^2 =0$.
	\item All (or some) error-error-covariances shall be zero: \newline  ${\forall}k,k'{\in}\left\{1,{\dots},m\right\}:\sum _{i=1}^n\epsilon _{i,k}\cdot \epsilon _{i,k'}=0$.
	\item All (or some) latent-error-covariances shall be zero: \newline ${\forall}q{\in}\left\{1,{\dots},Q\right\}:{\forall}k{\in}\left\{1,{\dots},m\right\}:\sum _{i=1}^n Z_{i,q}\cdot \epsilon_{i,k}=0$.
\end{itemize}  

These additional constraints may either be set strongly in the sense of constrained optimization or more softly by adding a penalty term  to $F$ like $P\cdot\left(\sum _{i=1}^n c_{i,k}\right)^2$ where $P>0$ is a penalty constant chosen big enough to enforce the constraint $c$ to the desired precision. In practice, it is often advisable  to follow the softer strategy because finite samples usually  violate these equations to some extent, too.

The minimizer of (\ref{objfun}) gives values for all parameters and also for all latent variables. Plugging these values back into equation (\ref{geq}) one gets estimates of all individual errors and from this error (co)variances can be calculated.

There are several strategies to choose the weights $w_l$ in the above objective functions. If one knew the error variances $\sigma(\epsilon_l)^2$,  it would be sensible to set $w_l:=1/\sigma(\epsilon_l)^2$ because this would turn multiplication of an equation with an arbitrary factor into an invariance operation. Moreover, it easy to see that for any $w_l:=c/\sigma(\epsilon_l)^2,c\in\mathbb{R}^+$ and under the additional assumption that the errors are independent and normally distributed, the minimizer of $F$ is the maximum-likelihood estimate because then for the normalized equations $g_l':=g_l/\sigma(\epsilon_l)$ all errors $\epsilon_l':=\epsilon_l/\sigma(\epsilon_l)$ are distributed by the standard normal distribution. Hence, they have the same variance and hence the probability density $f$ of a particular observation $A_{i,\cdot}$ is given by 
\begin{equation}
	f(A_{i,\cdot})\sim \prod_{l=1}^{m}exp\left(\frac{-g_l\left(A_{i,\cdot},\left\{Z_{i,q}\right\},\{p_s\}\right)^2}{2}\right)
\end{equation}	
and hence by standard arguments the least square minimizer maximizes likelihood (the argument is almost literally the
same calculation as in  \cite[p. 32]{SW1988}).

In general,  knowledge of the true $\sigma(\epsilon_l)$ will not be available and then the following strategies can be used:
\begin{itemize}
	\item Strategy $W_1$: For unweighted least square one simply chooses $w_l=1$. The objective function then  realizes a uniform least square approach,  and hence, just like $F_{ULS}$  in SEM, it requires equation defects to be measured on the same scale.
	\item Strategy $W_n$:  Assuming that the measured data is of good quality, one may assume that such equations that relate just one latent variable to the manifest data are more important while equations that relate many latent variables are of more hypothetical nature and thus might be less important. This motivates the choice  $w_l=\frac{1}{n^{L_l}}$,  where  $L_l$  is the number of latent variables in equation  $g_l$.
	\item Strategy $W_w$: This is a two-step-strategy: First use $W_n$ to get estimates of all variables (and thus of the error variances) and then use the reciprocal error variances as weights in a second step. 
	\item Strategy $W_o$: This is a self-consistency  adapting method: The weights $w$ and a proportionality factor $K$ are included into the parameters to be estimated. Hence, the new set of parameter is $\left\{p_s,w_l,K\right\}$ and there are additional constraints $\sum_l w_l=1, w_l\geq0,K>0$. The objective function is extended by a penalty term 
\begin{equation}
	F^{(o)}\left(\left\{Z_{i,q}\right\},\{p_s,w_l,K\}\right):=F_w\left(\left\{Z_{i,q}\right\},\{p_s\}\right)+P\cdot \sum_{l=1}^m \left(w_l-K/\epsilon_l^2\right)^2
\end{equation}	
Here $P>>0$ is some large penalty number.  Alternatively, the constraints $w_l=K\cdot\epsilon_l^2$  can be set exactly, e.g. by Lagrange multipliers. 
	\item Strategy $W_a$: This is an optimization algorithm that is based on the fact that the true inverse error variances and accordingly chosen weights are proportional, so that the angle between them is minimal.  Hence, one considers the function $H:\mathbb{R}^m\rightarrow\mathbb{R}, w\mapsto \frac{w\cdot \frac{1}{\sigma(w)^2}}{|w|\cdot |1/\sigma(w)^2|}$ where $\sigma(w)\in\mathbb{R}^m$ denotes the vector of standard deviations of (\ref{geq}) calculated with the minimizer of (\ref{objfun}) for weights $w$ and inverse of a vector is understood element-wise. Then a optimization algorithm is used to maximize $H$ with respect to the conditions $\sum_l w_l=1, w_l\geq0$. The maximizer is a weight vector that will then be used to minimize (\ref{objfun}) to give the final results. 
\end{itemize}

Strategy $W_1$ is theoretically sound if the assumptions are met. Strategies $W_n,W_w$ are heuristic ones that may give results biased to some extend. 

Method $W_a$ is theoretically sound: There is a unique maximizer of $H$ on the compact parameter set defined by the restrictions on $w$. This maximizer is given by the true inverse error variances and yields $H(w)=1$.  With these correct weights the method produces maximum likelihood estimates. A drawback is, of course, that the maximization of $H$ is extremely expensive in terms of computing power, because each evaluation of $H$ means to calculate a solution of (\ref{objfun}).   

Finally, $W_o$ is appealing as it should have the same theoretical properties as $W_a$ but with less computational demand. However, numerical minimization of $F^{(a)}$ is delicate: There is a global minimizer that sets all weights but one to zero and gives weight 1 to the equation with the smallest error. It is thus essential to start local minimization with a good initial guess (e.g. obtained from $W_w$).

Besides these methods there are some more. Especially one might think of iterating $W_w$  by taking each time the error estimates as a basis for the new weights. Theoretically, the correct weight vector is a fixed point of this iteration. Experience shows, however, the following behaviour: For a small number of iterations results are good but obviously the fixed point is not attractive and usually the iteration runs away from the correct solution and ends in the same  degenerate solution as described for $W_o$.   

Note that general principles of calculus  guarantee the existence of a minimizer  $\left\{\widehat
{{Z}_{i,q}}\right\},\{\widehat {p_s}\}$, i.e. values such that the function value of the objective function equals the infimum  over the total parameter space  $F_1\left(\left\{\widehat {{Z}_{i,q}}\right\},\{\widehat
{p_s}\}\right)={\mathit{inf}}F_1\left(\left\{{Z}_{i,q}\right\},\{p_s\}\right)$  if the parameter space is
compact and all the  $g_l$  are continuous. If there is an argument that the absolute values of latent
variables and parameters are bounded by some (maybe very large) numbers,  compactness is given. Hence, existence of
a minimizer is guaranteed in most situations. Uniqueness, however, is not guaranteed. If the functions  $g_l$ are twice
differentiable, it can be checked (Hesse matrix) if the minimizer is locally unique. However, it is a question of the
optimization method if other minimizers may be found. Thus, when applying the method, one thus should consider global methods as well. 

Yet another point is that of consistency. It seems that there are no general results on consistency this for this kind of problems. 
Regarding maximum likelihood estimation of nonlinear models, \cite{Philips1982}  has shown that the widespread belief that non-normality of errors leads always to non-consistency is false. However, in general, consistency cannot be assumed. Ivanov \cite[ch. 1]{Ivanov1997} gives a number of examples where nonlinear least square estimators are not consistent. Hence, theoretical investigations of consistency can only be expected for special models. In this paper only empirical evidence from simulation studies will be given.

Issues of the practical implementation of this approach in the Mathematica computer algebra system are given in \cite{Old2020}.    In the Mathematica implementation one can freely choose between the strategies described above, add additional constraints by hand or semi-automatic and choose between different  optimization algorithms. The full implementation is available on \url{https://myweb.rz.uni-augsburg.de/~oldenbre/sem/index.html}. 

The structural equations of linear SEM are a special case of CLSSEM.  The model equations  
${x}={\Lambda _x}\xi+\delta ,{y}={\Lambda_y}\eta +\epsilon ,\eta ={B\eta }+{\Gamma} \xi +\zeta $  
yield the following (unweighted) objective function
in vector notation (i.e.  ${x}$  is a vector of vectors etc. and ${x}^2:={x'x}$  and the sum over cases is implicit):
\begin{equation}\begin{split}
		F_1\left(\eta ,\xi ,{\Lambda _x,\Lambda _y,B,\Gamma }\right)=
		\delta^2+\epsilon^2+\zeta^2=\\
		\left({x}-{\Lambda_x}\xi \right)^2+\left({y}-{\Lambda _y}\eta\right)^2+\left({B}\eta+{\Gamma} \xi -\eta\right)^2=
		\\
		{x'x-x'\Lambda _x}\xi-\xi'{\Lambda_x}' x+{\Lambda_x}'\xi' \xi {\Lambda _x+y'y-y'\Lambda _y}\eta-\eta'{\Lambda_y'y}+\eta'{\Lambda _y' \Lambda _y}\eta+\\
		\eta'{B'B}\eta+\eta'{B}'{\Gamma}\xi-\eta'{B}'\eta    + \xi'{\Gamma'B}\eta  + \xi'{\Gamma' \Gamma} \xi -   \xi'{\Gamma'} \eta -\eta'{B}\eta-\eta'{\Gamma}\xi+\eta'\eta  
\end{split}\end{equation}
The optimization variables in this are the path coefficients from ${\Lambda_x,\Lambda_y,B,\Gamma}$ that are not preset from model specification and the values of all latent variables $\xi,\eta$ for all cases. 

It is rather obvious that it is not possible to generate closed form solutions for the system that states that the
gradient of this function is zero. Thus, numerical methods must be applied. However, it can be seen from this
expression that there is no need for  ${B-I}$  to be non-singular as it is in SEM.

\subsection{Some exact calculations}\label{sec:2calc}
It is evident that in general there is no closed form solution for the optimization problem specified above. In this subsection, however, exact solutions   for some very simple models are derived  and some insight can be gained from them.

A first example is the regression problem with error in both variables. Let $X,Y$ be centered, normally distributed random variables exactly related by $Y=a\cdot X, a\in\mathbb{R}$. These variables are measured with error as $x=X+\delta, y=Y+\epsilon, \delta\sim N(0,\sigma_\delta), \epsilon\sim N(0,\sigma_\delta)$.   As is well known \cite{Fuller1987}, standard linear regression will estimate $a$ biased toward zero. A practical solution (that is unfortunately not consistent) that improves the estimate in many situations  is to use orthogonal projection \cite{Glaister2001}. Modeling this in our approach with strategy $W_1$ yields the same solution: The model is $x=Z, y=a\cdot Z$ with a latent variable $Z$. This yields the objective function: $F(a,Z_1,\ldots,Z_n)=\sum_{i=1}^n \left[ (x_i-Z_i)^2+(y_i-a\cdot Z_i)^2\right]$.

Equating the partial derivatives to zero gives forall $i$: $0=\frac{\partial F}{\partial Z_i}=-2(x_i-Z_i)-2(y_i-aZ_i)\cdot a$. From this one gets $Z_i=\frac{x_i+a\cdot y_i}{1+a^2}$. 

The derivative with respect to $a$ gives: $0=\frac{\partial F}{\partial a}=-2\sum_{i=1}^n (y_i-a\cdot Z_i)\cdot Z_i$. 
Plugging in the above solutions for $Z_i$ and using the shortcuts $A=\sum_i x_i^2, B=\sum_i y_i^2, C=\sum_i x_i y_i$ yields, after  some algebra, the solution: $a=\frac{B-A+\sqrt{(A-B)^2+4C^2}}{2C}$and this is exactly the same solution as the one given by  \cite{Glaister2001}. 

For the regression example the strategy $W_a$ can also be worked out exactly using a Lagrange multiplier approach. The resulting equations are too lengthy to be displayed here. There are three solutions for $w$, namely $0,1$ and a third one somewhere in between.  The estimate of $a$ derived from this is, however, not unbiased, although it is better that the standard regression.  

A second example is that of a pure reflective measurement model with model equations $x_j=\lambda_j\cdot Z+\epsilon_j, j=1..m,\lambda_1=1$. This yields with strategy $W_1$: 

$F=\sum_{i=1}^n \sum_{j=1}^m (x_{j,i}-\lambda_j\cdot Z_i)^2$.  Equating the partial derivative for $Z_i$ to zero gives
$0=\frac{\partial F}{\partial Z_i}=-2\sum_{j=1}^m (x_{j,i}-\lambda_j\cdot Z_i)\cdot\lambda_j$. Thus: $Z_i=\frac{\sum_j x_{j,i}\cdot\lambda_j}{\sum_j \lambda_j}$. This shows that in this special case, we get a composite model. In general, with more than one interacting latent variables, this is not the case.

\subsection{A comment on Identification}\label{sec:2ident}
For the CLSSEM optimization problem to identify parameters, it is sufficient that partial derivatives vanish only on a discrete set. The above  regression example shows that there are only two critical points of $F$ (as the gradient vanishes only in two points) and only one of them is a minimum. Thus the solution is determined uniquely.  Hence, this model is identified in CLSSEM, but not in SEM (there are four parameters to be determined from the three numbers in the covariance matrix). This toy example shows that CLSSEM can be identified in cases were SEM is not.

\subsection{Fitting measures for CLSSEM}\label{sec:2fit}
If one has
reason to assume the  $\epsilon _l$  to be independent and normally distributed, then, of course, one can perform a 
$\chi ^2$  hypothesis test of model fit. 
If one assumes the  $\epsilon _{i,k}$  to be distributed according to  $N(0,\sigma ^2)$  then the least square minimizer of  $F_1$ is also a maximum likelihood estimation.

Without such distributional assumptions, an obvious measure for model fit is the mean of residuals, i.e.  $R:=\sqrt{\frac{F_1^{\mathit{min}}}{n\cdot m}}$ . 
This allows to compare different models if their equations have been defined in a comparable manner without any further assumptions about the distribution of the errors. It is not
possible to perform a hypothesis test if the distribution of  $F^{\mathit{min}}$  is not known. 
What can be done, however,  is to compare the fit to the fit of the null data, i.e. one randomly permutes the  $n$ 
entries of each of the  $k$  data vectors. The resulting minimum of the objective function can then be compared to the one
from the model-data fit.

\section{Case studies}\label{sec:3}
This paragraph presents studies of the performance of CLSSEM on various problems with simulated data. 

\subsection{A toy example: Regression }\label{sec:3bollen}
As a first almost trivial example I'll look at a linear regression model where the independent and dependent variable are measured with error by two manifest variables each. The model is therefore consists of a latent variable $X$ distributed by $N(0,1)$, a regression coefficient $a$ which was set to $0.5$ for the simulations and four observed variables $x_i=X+\delta_i,y_i=a\cdot X+\epsilon_i, i=1,2$ with $\sigma(\delta_1)= 0.5,\sigma(\delta_2)= 0.2,\sigma(\epsilon_1)= 0.2,\sigma(\epsilon_1)= 0.1$.  Result in table \ref{tab:reg} show that $W_1,W_n$ (which are identical in this case) might be slightly biased but that $W_w$ performs very well. $W_o,W_a$  have higher variance in the  estimates and $W_o$ may also be slightly biased, unlike $W_a$.

\begin{table}
	\caption{Simulation study for a regression model with four manifest variables: Errors of estimates of $a$ over 25 simulations)}
	{\footnotesize
		\begin{tabular}{ccccccc} \hline 
   $n$ & $a$ & $W_1$ & $W_n$ & $W_w$ & $W_o$ & $W_a$  \\ \hline 
100 & 0.5  & -0.029(0.012)  & -0.029(0.012)  & -0.008(0.012)  & -0.035(0.053)  & -0.023(0.022) \\ 
200 & 0.5  & -0.027(0.013)  & -0.027(0.013)  & -0.004(0.012)  & -0.044(0.041)  & -0.011(0.020) \\ 
500 & 0.5  & -0.027(0.007)  & -0.027(0.007)  & -0.005(0.007)  & -0.034(0.019) & -0.006(0.005)\\ 
			\hline
	\end{tabular}}
	\label{tab:reg}
\end{table}

\subsection{Bollen{}'s democracy data set }\label{sec:3bollen}
A classic example of a non-trivial SEM model is Bollen's model of democracy and industrialization \cite[p. 332]{Bollen1989}.  This model can be estimated perfectly with the standard SEM approach where ML estimation gives consistent unbiased estimates. Surprisingly, this model is the one that shows the strongest sensitivity for the choice of weights. While for many other models the uniform weight strategy $W_1$ gives good results, here this simple strategy will be disappointing, as we shall see.      

The model has three latent Variables  $\mathit{ind}60,\mathit{dem}60,\mathit{dem}65$  and eleven manifest variables 
$x_1,{\dots},x_3,y_1,{\dots},y_8$. The model equations are (with error variables $\delta,\epsilon,\gamma$ and intercept variables $s_i,t_i)$:

\begin{equation}\begin{split}
		x_1=1{\cdot}\mathit{ind}60+t_1+\delta_1,   x_2=c_2{\cdot}\mathit{ind}60+t_2+\delta_2, x_3=c_3{\cdot}\mathit{ind}60+t_3+\delta_3\\
		y_1=1{\cdot}\mathit{dem}60+s_1+\epsilon_1, y_2=d_2{\cdot}\mathit{dem}60+s_2+\epsilon_2\\
		y_3=d_3{\cdot}\mathit{dem}60+s_3+\epsilon_4, y_4=d_4{\cdot}\mathit{dem}60+s_4+\epsilon_4\\
		y_5=1{\cdot}\mathit{dem}65+s_5+\epsilon_5, y_6=d_6{\cdot}\mathit{dem}65+s_6+\epsilon_6\\
		y_7=d_7{\cdot}\mathit{dem}65+s_7+\epsilon_7, y_8=d_8{\cdot}\mathit{dem}65+s_8+\epsilon_8\\
		\mathit{dem}60=b_1{\cdot}\mathit{ind}60+\gamma_1,\mathit{dem}65=b_2{\cdot}\mathit{ind}60+b_3{\cdot}\mathit{dem}60+\gamma_2
	\end{split}
\end{equation}
Furthermore, Bollen assumed that (only) the some residual covariances may differ from 0 but the simulations in this study deviate from this by simulating so that all of them are independent.  
In the description of the algorithm, the notion  $N(m,\sigma ,n)$  is used
to denote a vector in  $\mathbb{R}^n$  with normally distributed, independent random numbers with given mean  $m$  and standard deviation  $\sigma $. The algorithm produces 
$$X_1,X_2,X_3,Y_1,Y_{2,}{\dots},Y_{8,}\mathit{ind}60,\mathit{dem}60,\mathit{dem}65{\in}\mathbb{R}^n$$  from the following
input: sample size  $n{\in}\mathbb{N}$,   parameters  $b_1,b_2,b_3,c_2,c_3,d_2,{\dots},d_4$, $d_6,{\dots},d_8 {\in} \mathbb{R}$,  variance parameters  $\sigma_{X1}, \sigma_{X2}, \sigma_{X3}, \sigma_{Y1}, \sigma_{Y2}, \sigma_{Y3}, \sigma_{Y4}, \sigma_{Y5}, \sigma_{Y6}$, $\sigma_{Y7}, \sigma_{Y8}, \sigma_1, \sigma_2{\in} \mathbb{R}$

The steps of this algorithm are:

\begin{enumerate}
	\item $\mathit{ind60}:=N(0,1,n)$
	\item $X_1:=1.0{\cdot}\mathit{ind60}+N\left(0,\sigma_{X1},n\right);
	X_2:=c_2{\cdot}\mathit{ind60}+N\left(0,\sigma_{X2},n\right);$\\
	$X_3:=c_3{\cdot}\mathit{ind60}+N(0,\sigma_{X3} ,n);$
	\item $\mathit{dem60}:=b_1{\cdot}\mathit{ind60}+N(0,\sigma_1 ,n)$
	\item  $Y_1:=1.0{\cdot}\mathit{dem60}+N\left(0,\sigma_{Y1} ,n\right);
	Y_2:=d_2{\cdot}\mathit{dem60}+N(0,\sigma_{Y2} ,n);$\\
	$Y_3:=d_3{\cdot}\mathit{dem60}+N\left(0,\sigma_{Y3} ,n\right);
	Y_4:=d_4{\cdot}\mathit{dem60}+N(0,\sigma_{Y4} ,n);$
	\item $\mathit{dem65}:=b_2{\cdot}\mathit{ind60}+b_3{\cdot}\mathit{dem60}+N(0,\sigma_2,n)$
	\item  $Y_5:=1.0{\cdot}\mathit{dem60}+N\left(0,\sigma_{Y5},n\right);
	Y_6:=d_6{\cdot}\mathit{dem60}+N(0,\sigma_{Y6} ,n);$\\
	$Y_7:=d_7{\cdot}\mathit{dem60}+N\left(0,\sigma_{Y7},n\right);
	Y_8:=d_8{\cdot}\mathit{dem60}+N(0,\sigma_{Y8} ,n);$
\end{enumerate}

The parameters where chosen for  the parameters are
$$(b_1,b_2,b_3,c_2,c_3,d_2,{\dots},d_4,d_6,{\dots},d_8) = (1.2, 0.5, 0.8,      0.7, 0.9,     0.3, 0.9, 1.7,     0.6, 0.4, 1.3)$$ 
and for the error standard deviations 
$$ (\sigma_{X1}, \sigma_{X2}, \sigma_{X3}, \sigma_{Y1}, \sigma_{Y2}, \sigma_{Y3}, \sigma_{Y4}, \sigma_{Y5}, \sigma_{Y6}, \sigma_{Y7}, \sigma_{Y8}, \sigma_1, \sigma_2) =$$
$$ (0.1, 0.2, 0.3,      0.2, 0.1, 0.2, 0.3, 0.2, 0.1, 0.2, 0.3,     0.5, 0.2) $$
This choice reflects the idea that the error variances should differ and be of moderate size.

A first insight gained from the simulations  was that all methods give good estimates for the loadings in the measurement model (i.e. $c_2,c_3,d_2,{\dots},d_4$, $d_6,{\dots},d_8$ are estimated without noticeable error) but parameters in the inner structural model are subject to larger deviations. Thus, in the following  only these parameter estimates are considered.  
Table \ref{tab:6} records average difference between estimates and true values together with standard deviations of these errors in parentheses.   For each simulated sample  size $n$ a set of 25 simulations were conducted. Due to long run times method $W_a$ was performed only for small $n$.  

On this example $W_1$ shows strong bias, especially for parameter $b_2$. Here, $W_n$ performs very well with the weight estimating methods $W_w$ and $W_o$ with larger variances but the conjecture that they produce unbiased results can be held. Method $W_a$ performs very poorly. This is partly due to convergence problems.

\begin{table}
	\caption{Comparison of estimation methods on the democracy model: mean error and std deviation  over 25 simulations }
	{\scriptsize\begin{tabular}{cccccccc}  \hline
$n$ & var      & true value &   $W_1$ & $W_n$  &$W_w$ & $W_o$ &$W_a$\\ 	
100 & $b_1$  & 1.2  & 0.011(0.065)  & -0.012(0.057)  & -0.024(0.057)  & 0.012(0.063) & 0.032(0.064) \\ 
100 & $b_2$  & 0.5  & 0.238(0.188)  & 0.004(0.058)  & 0.052(0.071)  & 0.059(0.176) & 0.337(0.197)\\ 
100 & $b_3$  & 0.8  & -0.166(0.143)  & -0.006(0.044)  & -0.039(0.052) & -0.023(0.143) & -0.285(0.134)  \\ 
200 & $b_1$  & 1.2  & 0.012(0.039)  & -0.011(0.033)  & -0.022(0.034)  & 0.013(0.039) \\ 
200 & $b_2$  & 0.5  & 0.161(0.122)  & -0.02(0.051)  & 0.03(0.06)  & -0.019(0.138) \\ 
200 & $b_3$  & 0.8  & -0.104(0.104)  & 0.013(0.046)  & -0.021(0.054)  \
& 0.041(0.116) \\ 
500 & $b_1$  & 1.2  & 0.011(0.027)  & -0.012(0.024)  & -0.024(0.025)  \
& 0.022(0.03) \\ 
500 & $b_2$  & 0.5  & 0.237(0.082)  & 0.011(0.029)  & 0.061(0.038)  & \
0.066(0.077) \\ 
500 & $b_ 3$  & 0.8  & -0.165(0.068)  & -0.009(0.025)  & \
-0.043(0.032)  & -0.016(0.072) \\ \hline
	\end{tabular}}
	\label{tab:1}
\end{table}

\subsection{Ganzach{}'s nonlinear model }\label{sec:3ganzach}

To test CLSSEM performance on a nonlinear model I use  Ganzach's model as given by \cite{KB2009}. 

This model has three latent variables $\xi_1,\xi_2,\eta$ and nine indicator variables $x_1,...,x_6,y_1,y_2,y_3$. The model equations are $\eta=\gamma_1 \xi_1+\gamma_2 \xi_2+\omega_{11} \xi_1^2+\omega_{12} \xi_1 \xi_2+\omega_{22} \xi_2^2+O_\eta+\epsilon_0$ and $\forall i \in \lbrace 1,...,6 \rbrace : x_i=c_i\cdot \xi_{\lceil i/3 \rceil}+O_i+\epsilon_i, \forall i \in \lbrace 1,...,3 \rbrace : y_i=d_i\cdot\eta+O_{6+i}+\epsilon_{6+i}$ with $c_1=c_4=d_1=1$ and with error variables $\epsilon$ and linear offset variables $O$.

To illustrate what a user needs to do to estimate the model using the implementation in Mathematica, here is the complete command:  
{\scriptsize \begin{verbatim} mySEM[dataMatrix, {x1, x2, x3, x4, x5, x6, y1, y2, y3}, {eta, xi1, xi2}, 
    	{ {y1 == 1*eta + O8, e7}, {y2 == d2*eta + O8, e8}, {y3 == d3*eta + O9, e9},
    	  {x1 == 1*xi1 + O1, e1}, {x2 == c2*xi1 + O2, e2}, {x3 == c3*xi1 + O3, e3}, 
		       {x4 == 1*xi2 + O4, e4}, {x5 == c5*xi2 + O5, e5}, {x6 == c6*xi2 + O6, e6}, 
	     	{eta == gamma1*xi1 + gamma2*xi2 + om11*xi1^2 +	om12*xi1*xi2 + om22*xi2^2 + Oeta, e0 } } ] 
\end{verbatim} }

The generation of the sample data sets is done by the following algorithm:

\begin{enumerate}
	\item $\xi_1,\xi_2$ are sampled normally distributed with mean 0 and covariance matrix $\left(\begin{array}{cc}
		1 & 0.3 \\
		0.3 & 1
	\end{array}\right)$
	\item $\eta:=\gamma_1 \xi_1+\gamma_2 \xi_2+\omega_{11} \xi_1^2+\omega_{12} \xi_1 \xi_2+\omega_{22} \xi_2^2+N(0,0.3,n)$ with $\gamma_1=0.3,\gamma_2=0.2,\omega_{11}=0.5,\omega_{12}=0.3,\omega_{22}=0.2$
	\item $i\in\{1,2\}: y_i:=d_i \eta+N(0,0.1,n), y_3:=d_3 \eta+N(0,0.3,n)$ with $d=(1,0.8,1.3)$
	\item $i\in\{1,2\}: x_i:=c_i \xi_1+N(0,0.1,n),x_3:=c_3 \xi_1+N(0,0.3,n)$
	\item $i\in\{4,5\}: x_i:=c_i \xi_2+N(0,0.1,n),  x_6:=c_6 \xi_2+N(0,0.3,n)$ with $c=(1,0.7,1.2,1,0.5,0.9),d=(1,0.8,1.3)$.
\end{enumerate}

Note that the error variances differ to some extent.

The most difficult parameters to estimate are those in the inner, nonlinear model, hence in the following analysis concentrates on these variables. The table reports the  standard deviation of the differences between the vector of this correct values and the vector of the estimates for these variables:  $\gamma_1=0.3,\gamma_2=0.2,\omega_{11}=0.5,\omega_{12}=0.3,\omega_{22}=0.2$

Table \ref{tab:6} shows the comparison of estimates for this model. In this example  there were only minimal differences between the weighted and unweighted approach. 

\begin{table}
	\caption{Comparison of strategies on Ganzach’s model on simulated data (25 samples for each $n$)}
	{\scriptsize\begin{tabular}{cccccccc} \hline
	n &variable & true              & $W_1$  & $W_n$  & $W_w$  & $W_o$  & $W_a$  \\ \hline
100 & $\omega_{11}$  & 0.5  & -0.006(0.021)  & -0.009(0.021)  & -0.009(0.023)  & 0.017(0.024)  & -0.025(0.021) \\ 
100 & $\omega_{12}$  & 0.3  & -0.013(0.038)  & -0.015(0.039)  & -0.008(0.034)  & -0.009(0.043)  & -0.044(0.051) \\ 
100 & $\omega_{22}$  & 0.2  & -0.007(0.036)  & -0.012(0.032)  & -0.003(0.03)  & -0.017(0.038)  & 0.023(0.019) \\ 
100 & $\gamma_{1}$  & 0.3  & -0.017(0.117)  & -0.017(0.115)  & -0.018(0.114)  & -0.008(0.121)  & 0.039(0.064) \\ 
100 & $\gamma_{2}$  & 0.2  & -0.016(0.057)  & -0.016(0.052)  & -0.009(0.054)  & -0.022(0.059)  & 0.017(0.065) \\
500 & $\omega_{11}$  & 0.5  & -0.006(0.021)  & -0.009(0.021)  & -0.009(0.023)  & 0.017(0.024)  & -0.019(0.024) \\ 
500 & $\omega_{12}$  & 0.3  & -0.013(0.038)  & -0.015(0.039)  & -0.008(0.034)  & -0.009(0.043)  & -0.009(0.039) \\ 
500 & $\omega_{22}$  & 0.2  & -0.007(0.036)  & -0.012(0.032)  & -0.003(0.03)  & -0.017(0.038)  & -0.012(0.022) \\ 
500 & $\gamma_{1}$  & 0.3  & -0.017(0.117)  & -0.017(0.115)  & -0.018(0.114)  & -0.008(0.121)  & 0.026(0.069) \\ 
500 & $\gamma_{2}$  & 0.2  & -0.016(0.057)  & -0.016(0.052)  & -0.009(0.054)  & -0.022(0.059)  & -0.001(0.033) \\ \hline
	\end{tabular}}
	\label{tab:6}
\end{table}

\subsection{Muthen{}'s nonlinear model }\label{sec:3muthen}

A similar quadratic model is that by  \cite{AM2019}. 
The generation of the sample data sets is done by the following algorithm that tries to mimic the data generation in their publication:

\begin{enumerate}   
	\item $\eta_1,\eta_2$ are sampled normally distributed with mean 0 and covariance matrix $\left(\begin{array}{cc}
		1.2 & 0.4 \\
		0.4 & 0.8
	\end{array}\right)$
	\item $\eta_3:=B_1\cdot\eta_1 + B_2\cdot\eta_2 + B_3\cdot\eta_1 \cdot\eta_2+N(0,0.2,n),
	\eta_4:=B_4\cdot\eta_3+N(0,0.1,n) $ with $B_1 = 0.1,  B_2 = 0.3, B_3 = 0.2, B_4 = 0.7$
	\item $i\in\{1,..,12\}: y_i:=c_i \cdot\eta_{\lceil i/3\rceil}+N(0,0.1\cdot(1+(i \mod 3)),n)$ with $c=(1,0.5,0.7,1,0.7,0.4,1,1.2,0.4,1,0.8,0.9)$
\end{enumerate}

Just with the model above, the most difficult parameters to estimate are those in the inner, nonlinear model. 
Table \ref{tab:7} shows mean estimate errors and standard deviations for 25 simulations with $n=100$, resp. $n=500$ sample sizes (for $W_a$ only 10 simulations were performed due to long run times).   The results are fairly good. 

\begin{table}
	\caption{Asparouhov \& Muthén model (25 resp 10 imulations for each $n$)}
	{\scriptsize
\begin{tabular}{cccccccc} \hline 
   $n$ & Variable & true & $W_1$             & $W_n$             & $W_w$          & $W_o$    & $W_a$  \\ \hline 
   100 & $B_1$ & 0.1    & -0.015(0.033)  & -0.004(0.034)  & -0.013(0.033) & -0.023(0.031) & -0.019(0.030) \\ 
100 & $B_2$   & 0.3  & -0.035(0.031)  & -0.015(0.028)  & -0.033(0.027)  & -0.070(0.033) &   -0.002(0.040)\\ 
100 & $B_3$  & 0.2  & -0.015(0.033)  & -0.002(0.033)  & -0.018(0.032)  & -0.034(0.036)  & 0.004(0.024)\\ 
100 & $B_4$  & 0.7   & -0.084(0.054)  & -0.115(0.038)  & -0.026(0.131) & -0.085(0.05)  &  -0.088(0.056)\\ 
500 & $B_1$  & 0.1   & -0.017(0.013)  & -0.007(0.014)  & -0.013(0.014) & -0.020(0.034)  & 0.001(0.011)  \\ 
500 & $B_2$  & 0.3  & -0.040(0.017)  & -0.017(0.015)  & -0.036(0.017)  & -0.066(0.037) & 0.004(0.025) \\ 
500 & $B_3$  & 0.2  & -0.025(0.009)  & -0.011(0.008)  & -0.023(0.008) & -0.040(0.033) & 0.006(0.016) \\ 
500 & $B_4$  & 0.7  & -0.085(0.035)  & -0.12(0.024)  & 0.006(0.048)  & -0.082(0.029) & -0.048(0.030) \\ 
			\hline
	\end{tabular}}
	\label{tab:7}
\end{table}

\subsection{A model with exponential relations }\label{sec:3exp}

This section shows that even exponential relations between latent variables are easily modeled.

\begin{enumerate}   
	\item $X_0:=N(0,0.1,n),Y_0:=3\cdot \exp(X_0/2)$ 
	\item $x_1:=X_0+N(0,0.1,n), x_2:=c_2\cdot X_0+N(0,0.2,n), c2:=0.7$ 
	\item $y_1:=Y_0+N(0,0.2,n), y_2:=d_2\cdot Y_0+N(0,0.1,n), d2:=0.9$ 
\end{enumerate}

The model equations thus are $x_1 = X_0, x_2 = c_2\cdot X_0,y_1 = d_1\cdot \exp(X_0\cdot k_1),y2 = d_2\cdot d_1\cdot \exp(X_0\cdot k_1)$. 
The results of 25 simulations for two different sample sizes are collected in table \ref{tab:8exp}. 
In this example, there is virtually no difference between all  strategies. 

\begin{table}
	\caption{Exponential model (25  simulations for $n$)~~~~~~~~~~~~~~~~~~~~~~}
	{\scriptsize\begin{tabular}{cccccccc} \hline 
			$n$ & Variable & true & $W_1$             & $W_n$             & $W_w$          & $W_o$    & $W_a$  \\ \hline 
100 & $c_2$  & 0.7  & -0.008(0.029)  & -0.008(0.029)  & -0.001(0.03)  & -0.007(0.044)  & -0.018(0.041) \\ 
100 & $d_1$  & 3  & 0.082(0.151)  & 0.082(0.151)  & 0.081(0.152)  & 0.082(0.148)  & 0.088(0.15) \\ 
100 & $d_2$  & 0.9  & -0.003(0.004)  & -0.003(0.004)  & -0.003(0.004) & -0.005(0.008)  & -0.003(0.004) \\
100 & $k_1$  & 0.5  & -0.007(0.012)  & -0.007(0.012)  & -0.001(0.013)  & -0.016(0.024)  & -0.016(0.018) \\ 
200 & $c_2$  & 0.7  & -0.019(0.02)  & -0.019(0.02)  & -0.013(0.02)  & -0.005(0.058) \\ 
200 & $d_1$  & 3  & 0.011(0.143)  & 0.011(0.143)  & 0.011(0.143)  & 0.013(0.141) \\ 
200 & $d_2$  & 0.9  & 0.(0.004)  & 0.(0.004)  & 0.(0.004)  & 0.004(0.012) \\ 
200 & $k_1$  & 0.5  & -0.008(0.014)  & -0.008(0.014)  & -0.003(0.013)  & 0.002(0.042) \\ 
500 & $c_2$  & 0.7  & -0.008(0.012)  & -0.008(0.012)  & -0.002(0.013)  & -0.008(0.022) \\ 
500 & $d_1$  & 3  & -0.007(0.073)  & -0.007(0.073)  & -0.008(0.073)  & -0.004(0.07) \\ 
500 & $d_2$  & 0.9  & 0.(0.003)  & 0.(0.003)  & 0.(0.003)  &  0.003(0.009) \\ 
500 & $k_1$  & 0.5  & -0.007(0.006)  & -0.007(0.006)  & -0.001(0.007)  & 0.003(0.042) \\ 
			\hline
	\end{tabular}}
	\label{tab:8exp}
\end{table}

\subsection{A model with implications }\label{sec:3impl}

Correlations capture no directional information between variables. To retrieve directional information other means such as implicative influence.  Implications between two centered, numerical variables $x\Rightarrow y$ are not affected by negative values of $x$, but for positive values of $x$, one expects positive values of $y$ as well. There is a branch of statistics that is about implications and seems to be rather detached from the mainstream \cite{Gras2008}. My suggestion is to incorporate implications using the function $\theta(x):= (x+|x|)/2$. Using this as a linking function means that the path coefficient measures implication, not correlation. The following simulation describes a very simple application and it is only meant to illustrate the principal usability of such models. A systematic study will be given in another publication.

\begin{enumerate}   
	\item $X_0:=N(0,1,n) $ 
	\item $x_1:=X_0+N(0,0.3,n), x_2:=0.7\cdot X_0+N(0,0.15,n)$ 
	\item $y:=0.4\cdot X0+0.8\cdot \theta(X_0)+N(0,0.2,n)$
\end{enumerate}

The results of 25 simulations for two different sample sizes are collected in table \ref{tab:8}. All strategies show good results, with $W_w$ being best and "$W_o$ last, but still acceptable. Due to long runtimes $W_a$ was not evaluated.  
\begin{table}
	\caption{Implicative model~~~~~~~~~~~~~~~~~~~~~~}
	{\footnotesize\begin{tabular}{ccccccc} \hline
			n &variable & true value  & $W_1$   & $W_n$ & $W_w$ & $ W_o$   \\ 
100 & $c_2$  & 0.7  & -0.014(0.021)  & -0.014(0.021)  & 0.000(0.022)  & -0.017(0.021) \\ 
100 & $d_1$  & 0.4  & -0.009(0.032)  & -0.009(0.032)  & 0.001(0.036)  & -0.011(0.031) \\ 
100 & $d_2$  & 0.8  & -0.020(0.129)  & -0.020(0.129)  & 0.001(0.142)  & -0.030(0.123) \\ 
			\hline
	\end{tabular}}
	\label{tab:8}
\end{table}

\subsection{A real world example }\label{sec:3impl}
This section gives a small example of the kind of problems this method was developed for. Im mathematics education one is concerned with the algebraic competencies that students should develop. One ability in this domain is to do syntactical transformations such as expanding and simplifying expressions. This ability will be model by a latent variable $T\in[0,1]$. Another ability is to substitute values into expressions and replace sub-expressions. This ability will be modeled by $S\in[0,1]$. In this model $S\cdot T$ can be interpreted as a measure for having both abilities. Both $S,T$ are measured by a set of items $x_i=\lambda_i\cdot S+\delta_i,i=1..9, y_j=\mu_j\cdot T+\delta'_j,i=1..10$. Moreover, there are equations that model success in solving a task $I$ by a relation of the form  
$I=s\cdot S+t\cdot T+c\cdot I\cdot S+\epsilon$. The fitting procedure estimates ($W_w$, but other methods give very much the same results): $s=-0.65, t=0.00, c=1.9$. These findings indicate that in fact having both competencies are essential for a good performance on this task. 

\section{Conclusion}\label{sec:concl}

Summing up the results from the simulation studies, these are the key findings: The proposed methods gives good estimates in many models. Even those models that are sensitive to the choice of weights can usually be estimated by the proposed methods of self-consistent weight selection.  For nonlinear models, CLSSEM performs very well and models with exponential and implicative relations can be handled quite well.

This paper gives empirical evidence that CLSSEM works well for moderate sample sizes and allows to model with much more flexibility than existing approaches.  However, a lot of research questions are still open. On the technical side it desirable to have a more sophisticated handling of constraints that should cut down run-times. For practical applications one needs to develop reliable fit indices and techniques to allow hypothesis testing (extending the ideas sketched above). Moreover, the possibilities arising from using non-linear fit functions have to be evaluated: Especially piece-wise linear functions and implicative linkings  will be analyzed in follow-up research. Theoretical work should further elaborate questions of consistency for some special models. Thus, summing up, this paper provides some new insights and opens up new areas for research.

\end{document}